\documentclass[a4paper]{article}

\usepackage{INTERSPEECH2022}
\usepackage{tabularx}
\usepackage{kotex, comment, arydshln}
\usepackage{cite, url}
\usepackage{xcolor}
\usepackage{makecell}
\usepackage{multirow}
\usepackage[subtle]{savetrees}
\usepackage{hyperref}
\hypersetup{ 
    colorlinks=true,
    linkcolor=purple,
    filecolor=magenta,      
    urlcolor=purple,
}

\newcolumntype{C}{>{\small\centering\arraybackslash}X}

\title{Extended U-Net for Speaker Verification in Noisy Environments}
\name{Ju-ho Kim, Jungwoo Heo, Hye-jin Shim, and Ha-Jin  Yu$^\dag$\thanks{$^\dag$ Corresponding author}}
\address{School of Computer Science, University of Seoul, Republic of Korea}
\email{wngh1187@naver.com, jungwoo4021@gmail.com, shimhz6.6@gmail.com, hjyu@uos.ac.kr}

\begin{document}

\maketitle

\begin{abstract}
Background noise is a well-known factor that deteriorates the accuracy and reliability of speaker verification (SV) systems by blurring speech intelligibility. 
Various studies have used separate pretrained enhancement models as the front-end module of the SV system in noisy environments, and these methods effectively remove noises. 
However, the denoising process of independent enhancement models not tailored to the SV task can also distort the speaker information included in utterances. 
We argue that the enhancement network and speaker embedding extractor should be fully jointly trained for SV tasks under noisy conditions to alleviate this issue. 
Therefore, we proposed a U-Net-based integrated framework that simultaneously optimizes speaker identification and feature enhancement losses. 
Moreover, we analyzed the structural limitations of using U-Net directly for noise SV tasks and further proposed \textit{Extended U-Net} to reduce these drawbacks. 
We evaluated the models on the noise-synthesized VoxCeleb1 test set and VOiCES development set recorded in various noisy scenarios. 
The experimental results demonstrate that the U-Net-based fully joint training framework is more effective than the baseline, and the extended U-Net exhibited state-of-the-art performance versus the recently proposed compensation systems. 
\end{abstract}


\section{Introduction}
Recently, deep learning has been applied in various fields with tremendous success, and in speaker verification (SV) tasks, deep neural network (DNN)-based embedding learning methods \cite{variani2014deep, snyder2018x} exhibit satisfactory results compared to traditional schemes \cite{reynolds2000speaker, dehak2010front}. 
Most embedding extractors perform reliably in clean scenarios but suffer from performance degradation in noisy environments. 
It is challenging to delicately and accurately extract speaker information from noise-contaminated utterances, because the background noise undermines the intelligibility and quality of speech \cite{wolfel2009distant, cai2020within}. 

\begin{figure}[t]
  \begin{center}
    \centering
    \vspace{-0.2cm}
    \includegraphics[width=\linewidth]{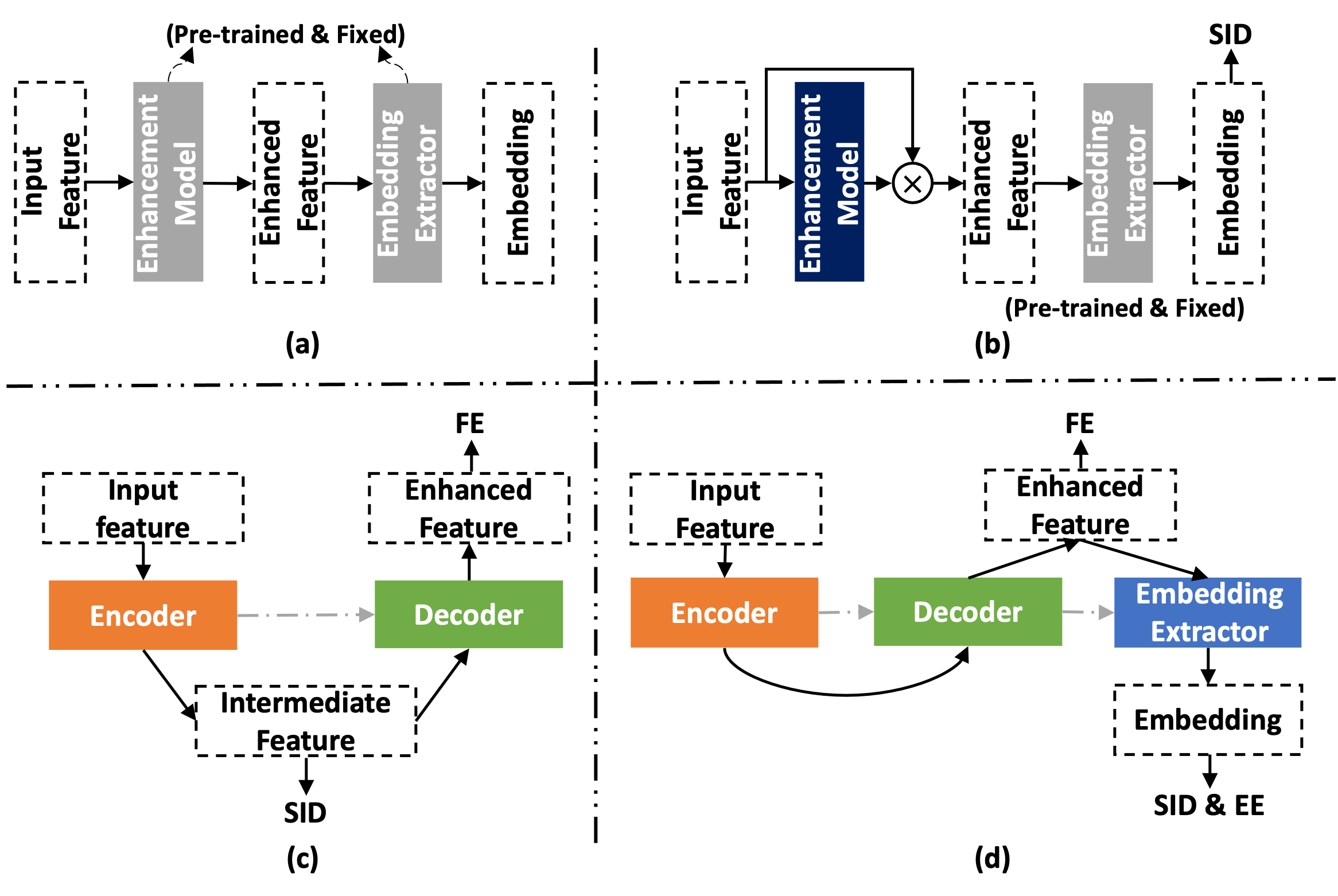} 
    \caption{Block diagrams of noisy speech compensation approaches for speaker verification tasks. 
    (a): Pretrained enhancement model, (b): VoiceID loss \cite{shon2019voiceid} (\textcircled{$\times$}\,denote element-wise multiplication), (c): U-Net-based system (Proposal 1), (d): Extended U-Net (Proposal 2)}
    \label{fig:diagrams}
  \end{center}
  \vspace{-0.4cm}
\end{figure}

Several studies have used a pretrained enhancement model as a front-end module for an embedding extractor (Figure \ref{fig:diagrams} (a)) to address this problem \cite{plchot2016audio, eskimez2018front, kolboek2016speech}. 
A denoising autoencoder, a traditional enhancement method, is trained to map noise speech to a clean counterpart using a loss function, such as the mean squared error (MSE) \cite{plchot2016audio, novotny2018use}. 
The denoising autoencoder can improve the auditory quality of noise speech, leading to better results for noisy SV scenarios. 
Nonetheless, the performance improvement for clean input is marginal, and even independent compensation processes not customized to downstream tasks may distort speaker information \cite{shon2019voiceid, kolboek2016speech}. 

VoiceID loss \cite{shon2019voiceid} was proposed to construct an enhancement model specialized for SV tasks. 
As presented in Figure \ref{fig:diagrams} (b), the mask output from the enhancement model is multiplied by the input feature element-wise to filter out redundant information polluted by noise. 
Then, the refined feature is fed to the pretrained SV model, and only the weights of the enhancement model are updated to reduce the speaker identification (SID) loss. 
This framework can generate an enhancement model suitable for SV but cannot jointly optimize both models. 
Therefore, the fixed SV model cannot continuously learn the modified distributions of enhanced features, which may lead to information mismatch \cite{wu2021joint}. 

To prevent the distortion of speaker information and mitigate information discrepancy caused by independent enhancement, we argue that the embedding extractor and enhancement model should be trained in a fully joint, rather than exclusive, manner. 
The U-Net \cite{ronneberger2015u} is an image-to-image translation network proposed in the biomedical image segmentation field based on the encoder-decoder structure, which delivers a processed output with the identical (or analogous) size of the input data. 
Due to the network's structural nature, the U-Net has been widely used in speech processing tasks, such as voice activity detection \cite{gusev2020deep} or speech enhancement \cite{giri2019attention}. 
Inspired by the architectural characteristics of the U-Net and its successful adaptation for the speech domain, we propose to exploit the U‑Net framework to optimize SID and feature enhancement (FE) losses simultaneously. 
The proposed U-Net-based system is depicted in Figure \ref{fig:diagrams} (c). 
This model is jointly trained to classify the speaker identity of the embeddings transformed from the intermediate features while reducing the Euclidean distance between the enhanced features from the decoder and its clean versions. 
Thus, the U-Net-based system can directly derive a noise-robust speaker embedding through this training process.

Although we devised a joint-trainable framework, the proposed U-Net-based system has several structural limitations. 
First, this system did not use enhanced features from the decoder to extract speaker embeddings. 
Furthermore, noise compensation is performed only with the encoder in the evaluation phase without using a decoder. 
Therefore, we hypothesize that the vanilla U-Net is not an ideal fully joint training structure for noise SV tasks and further propose an extended U-Net (ExU-Net) to improve these shortcomings. 
The ExU-Net (Figure \ref{fig:diagrams} (d)) is jointly trained by combining an additional embedding extractor with the U-Net framework. 
This extended structure can extract embeddings using the enhanced features and entirely exploit the jointly trained model for the evaluation. 
Moreover, we applied a metric learning loss for direct embedding enhancement (EE). 

The models were trained using VoxCeleb 1 \cite{voxceleb} training data and the MUSAN corpus \cite{snyder2015musan}. 
For evaluation in noisy environments, we used the VoxCeleb 1 test set synthesized from MUSAN and the VOiCES \cite{richey18_interspeech} development dataset containing various noise sources. 
The experiment results demonstrated that the U-Net-based system achieved improved performance compared to the baseline and that the ExU-Net exhibited state-of-the-art results for noisy scenarios. 

\section{Related work}
\label{sec:related work}
The SV studies for noisy environments have been conducted from various perspectives. 
Several researchers have focused on algorithmic methods that compensate for noisy utterances directly at the signal level \cite{borgstrom2012linear, movsner2018dereverberation}. 
At the feature level, data augmentation \cite{cai2020within} and feature normalization \cite{pelecanos2001feature} techniques were used to construct noise-robust SV systems. 
The SV research for noisy utterances has also been studied at the model level. 
Using a separate enhancement model as a front-end or back-end system for the embedding extractor can prevent performance deterioration due to noise \cite{shon2019voiceid, jung2020selective}. 
Furthermore, Cai \textit{et al.} \cite{cai2020within} proposed specialized loss functions that induce noise-mitigated speaker embeddings, and Wu \textit{et al.} \cite{wu2021joint} introduced asynchronous subregion optimization to avoid collision between losses. 
The present study also explored improving the performance of the noisy SV task at the model level and focused on a fully joint training scheme to alleviate the distortion of speaker information that may occur during noise compensation. 

\begin{figure}[t]
    \centering
    \includegraphics[width=1\linewidth]{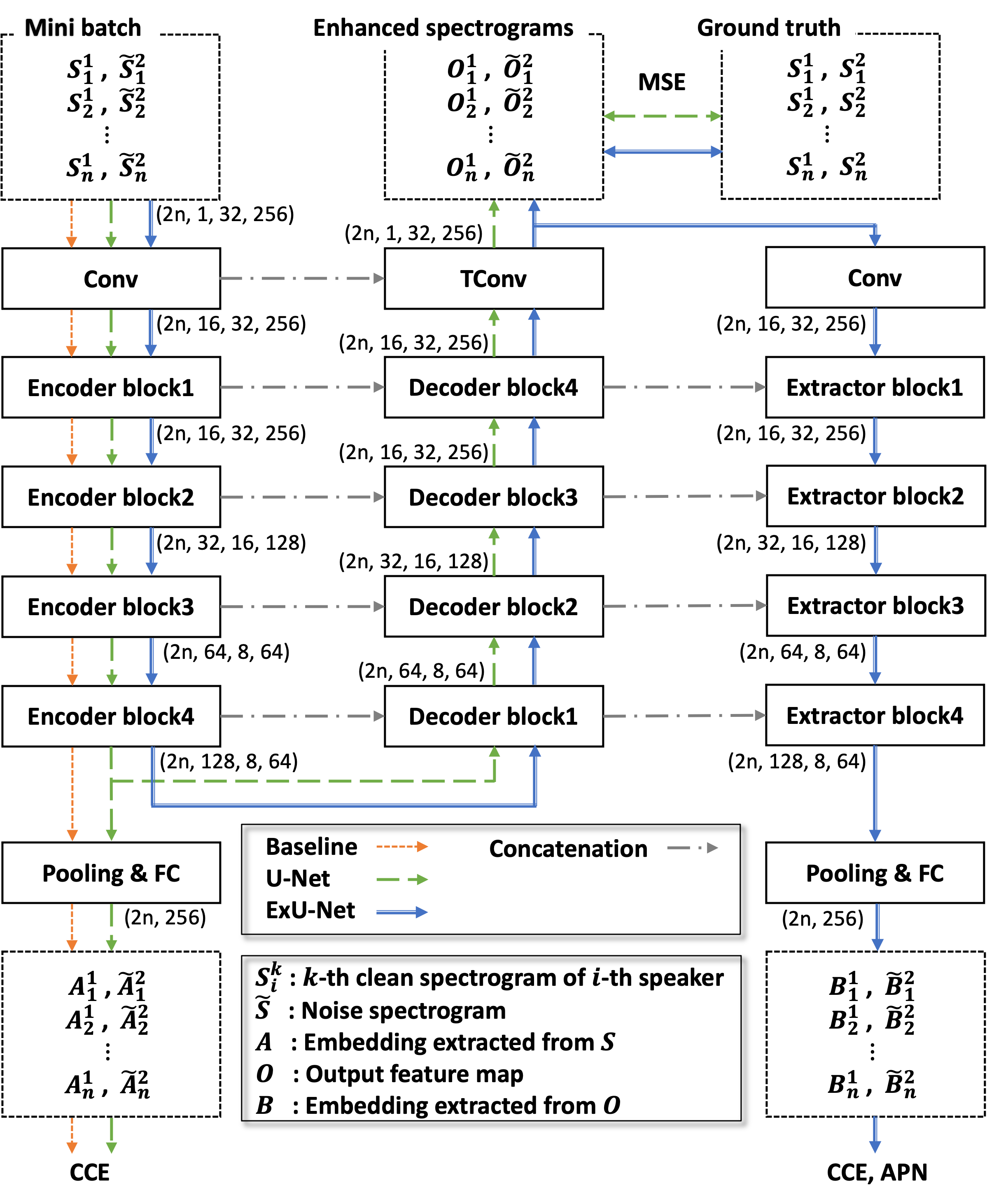}
    \vspace{-0.3cm}
\caption{
The training pipelines for the baseline, U-Net, and ExU-Net are indicated by orange, green, and blue arrows. 
The baseline is based on the encoder of the U-Net. 
The ExU-Net uses the enhanced spectrograms to extract speaker embedding through an additional extractor. 
The numbers in parentheses refer to the number of batches, channels, frequency bins, and frames of the output feature map. 
}
\label{fig:overall}
\vspace{-0.2cm}
\end{figure}

\section{Proposed frameworks}
\label{sec:proposed framework}
To mitigate the adverse effects caused by independent noise compensation, we speculate that the SV system should be trained by simultaneously optimizing the SID and FE losses. 
This section explicates the proposed fully joint training framework. 
As illustrated in Figure \ref{fig:overall}, we describe the models according to the scope of the framework in the following order: baseline, U-Net, and ExU-Net. 

\subsection{Baseline}
The encoder of our proposed frameworks captures latent features including speaker information from the input data. 
The residual network (ResNet) \cite{he2016deep} is one of the most successful DNNs in the past decade and has demonstrated outstanding performance as a speaker embedding extractor in SV tasks \cite{zhou2019deep}. 
Therefore, we used a ResNet-based system \cite{chung2020defence} as the encoder and designated it as the baseline for the direct comparison with the proposed models. 
The structure of the baseline is described in the left two columns of Table \ref{tab:structures}. 

The dotted thin orange arrows in Figure \ref{fig:overall} denote the training path of the baseline. 
We used a mini-batch ($\mathcal{M}$) consisting of clean and noise spectrograms to train the distributions of both data directly.
\begin{equation}
\begin{aligned}
    \mathcal{M} = &[\ S_{1}^{1}, S_{2}^{1}, ..., S_{n}^{1},&\\ 
                  & \ \ \tilde{S}_{1}^{2}, \tilde{S}_{2}^{2}, ..., \tilde{S}_{n}^{2}\ ] , &
\end{aligned}
\end{equation}
where $S_{i}^{j}$ indicates the $j$th clean spectrogram of the $i$th speaker, $\tilde{S}$ refers to the noise-synthesized spectrogram, and $n$ is the number of speakers for a single mini-batch. 
The input is processed into speaker embeddings, $A$, $\tilde{A} \in \mathbb{R}^{256}$, which are trained to classify the speaker identity via a categorical cross-entropy (CCE) criterion. 
\begin{equation}
\label{eq2}
\begin{aligned}
    A = g(f_{enc}(S)), \\ 
    \tilde{A} = g(f_{enc}(\tilde{S})),  
\end{aligned}
\end{equation}
where $f_{enc}$ indicates the encoder, and $g$ represents the final pooling and fully connected (FC) layers. 
Therefore, the loss function of the baseline is: 
\begin{equation}
\label{eq3}
    \mathcal{L}_{baseline} = \mathcal{L}_{CCE}.
\end{equation}

\subsection{U-Net-based system}
This study emphasized the significance of joint learning between the enhancement model and embedding extractor under noisy SV scenarios. 
Motivated by the speech enhancement research that successfully leveraged the structural characteristics of U-Net, we devised a fully joint training framework for noisy SV tasks based on the U-Net. 

The dotted thick green arrows in Figure \ref{fig:overall} illustrate the training process of the proposed U-Net-based system, and the structural details are specified in Table \ref{tab:structures}. 
The speaker embeddings are extracted with the identical process as the baseline and are trained with the SID task, as in Eq. (\ref{eq2}). 
The decoder network, $f_{dec}$, derives enhanced spectrograms ($O$ and $\tilde{O}$) of the same size as the input. 
The decoder uses the output of each encoder block to decode enhanced spectrograms with minimal information loss (dotted gray arrows in Figure \ref{fig:overall}). 
\begin{equation}
\begin{aligned}
    O = f_{dec}(f_{enc}(S)), \\ 
    \tilde{O} = f_{dec}(f_{enc}(\tilde{S})),  
\end{aligned}
\end{equation}
The enhanced spectrograms are optimized to reduce the $L$2 distance from the clean spectrograms of the input (i.\,e., the ground truth) using the MSE loss. 
\begin{equation}
    \mathcal{L}_{MSE} =  \frac{1}{2n}\sum^n_{i=1} (||{O}_{i}^{1} - {S}_{i}^{1}||^2_2 + ||\tilde{O}_{i}^{2} - S_{i}^{2}||^2_2), 
\end{equation}
The output of clean input is also mapped to the ground truth itself to prevent performance degradation when a clean utterance is input into the compensation system, as reported in \cite{jung2020selective}. 

In summary, the U-Net-based system is jointly trained to discriminate speakers and compensate for noisy utterances by simultaneously optimizing the following two losses: 
\begin{equation}
    \mathcal{L}_{Unet} = \mathcal{L}_{CCE} + \mathcal{L}_{MSE},
\end{equation}
Through this learning strategy, the proposed model can extract noise-robust speaker embeddings. 

\begin{table}[]
        \caption{Architectures of the proposed framework components. 
        Each decoder block (DB) includes a corresponding encoder block (EB) in which the number of input/output channels is reversed. 
        For the convolutional layer (Conv), the numbers inside parentheses indicate the kernel length, padding, and stride sizes, in turn. 
        (SE: squeeze-and-excitation module \cite{hu2018squeeze}, ASP: attentive statistics pooling \cite{okabe2018attentive}, TConv: transposed convolution layer)
        }
    \vspace{-0.1cm}
    \label{tab:structures}
\resizebox{\linewidth}{!}{%
    \begin{tabular}{c c | c c}
    \Xhline{2\arrayrulewidth}
    \textbf{Layer} & \textbf{Structure} & \textbf{Layer} & \textbf{Structure}\\
    \Xhline{2\arrayrulewidth}
        Conv & 
        $\begin{tabular}{c}
            Conv2d(7, 3, 2$\times$1)\\
            \end{tabular}$ & 
        DB1 &
        $\begin{tabular}{c}
            Eb4\\
            Concatenation \\
            Conv2d(1, 1, 1)\\
            \end{tabular}$
        \\
    \midrule
        \begin{tabular}{c}
        EB1 \\
        \end{tabular}&
        $\left [
            \begin{tabular}{c}
            Conv2d(3, 1, 1)\\
            Conv2d(3, 1, 1)\\
            SE\\
            \end{tabular}
        \right ]\times 3$ &
        DB2 &
        $\begin{tabular}{c}
            Eb3\\
            Concatenation \\
            TConv2d(2, 1, 2)\\
            \end{tabular}$
        \\
    \midrule
        \begin{tabular}{c}
        EB2 \\
        \end{tabular}&
        $\left [
            \begin{tabular}{c}
            Conv2d(3, 1, 2)\\
            Conv2d(3, 1, 1)\\
            SE\\
            \end{tabular}
        \right ]\times 4$ &
        DB3 &
        $\begin{tabular}{c}
            Eb2\\
            Concatenation \\
            TConv2d(2, 1, 2)\\
            \end{tabular}$
        \\
    \midrule
        \begin{tabular}{c}
        EB3 \\
        \end{tabular}&
        $\left [
            \begin{tabular}{c}
            Conv2d(3, 1, 2)\\
            Conv2d(3, 1, 1)\\
            SE\\
            \end{tabular}
        \right ]\times 6$ &
        DB4 &
        $\begin{tabular}{c}
            Eb1\\
            Concatenation \\
            Conv2d(1, 1, 1)\\
            \end{tabular}$
        \\
    \midrule
        \begin{tabular}{c}
        EB4 \\
        \end{tabular}&
        $\left [
            \begin{tabular}{c}
            Conv2d(3, 1, 1)\\
            Conv2d(3, 1, 1)\\
            SE\\
            \end{tabular}
        \right ]\times 3$ &
        TConv &
        $\begin{tabular}{c}
            Concatenation\\
            TConv2d(2$\times$1, 1, 2$\times$1)\\
            \end{tabular}$
        \\
    \midrule
        \begin{tabular}{c}
        Pooling \\
        FC\\
        \end{tabular}&
        $\begin{tabular}{c}
            ASP\\
            FC(256)\\
            \end{tabular}$ & 
        \multicolumn{2}{c}{-}
        \\
    \Xhline{2\arrayrulewidth}
    \end{tabular}
    \vspace{-0.2cm}
}
\end{table}

\begin{table*}[t!]
\caption{
    Experiment results (EER \%, $C_{det}^{min}$) obtained on the VoxCeleb1 test set and the noise scenarios synthesized with the MUSAN corpus under various SNRs ( $^{\dagger}$ : drawn from \cite{wu2021joint}). 
}
\vspace{-0.1cm}
\centering
\label{table:sota}
\resizebox{\textwidth}{!}{
\begin{tabular}{c|c|m{1.6cm}|m{1.5cm}m{1.8cm}m{1.8cm}m{1.5cm}m{1.5cm}m{1.5cm}m{1.5cm}} \Xhline{2\arrayrulewidth}
\multicolumn{2}{c|}{Training dataset} &  \multicolumn{1}{m{1.5cm}|}{\hfil Original $D$} & \multicolumn{6}{m{11.1cm}}{\hfil Original and noise augmentation $D + D^N$}\\ \hline
Noise type & SNR &\hfil Baseline &\hfil Baseline &\hfil VoiceID \cite{shon2019voiceid} $^{\dagger}$ &\hfil Wu \textit{et. al.} \cite{wu2021joint} &\hfil U-Net  &\hfil Cai \textit{et. al.} \cite{cai2020within} & \hfil ExU-Net-L & \hfil ExU-Net \\ \Xhline{2\arrayrulewidth}
\multicolumn{2}{c|}{\# Parameters}              &\hfil 1.39M        &\hfil 1.39M        &\hfil -            &\hfil -            &\hfil 3.41M        &\hfil -            &\hfil \textbf{1.38M}   &\hfil 4.81M \\ \hline
\multicolumn{2}{c|}{Original test set $\tau$}   &\hfil 3.75         &\hfil 3.69         &\hfil 6.79         &\hfil 7.6          &\hfil 3.57         &\hfil 3.12         &\hfil 3.23             &\hfil \textbf{2.76} \\\hline
\multirow{5}{*}{Babble} &\hfil 0                &\hfil 27.73        &\hfil 12.93        &\hfil 37.96        &\hfil 20.11        &\hfil 11.31        &\hfil 11.78        &\hfil 10.9             &\hfil \textbf{9.57} \\ \cline{2-10} 
                        &\hfil 5                &\hfil 14.54    &\hfil 7.23           &\hfil 27.12  &\hfil 12.02  &\hfil 6.62                &\hfil 5.97      &\hfil      6.04     &\hfil \textbf{5.52} \\ \cline{2-10} 
                        &\hfil 10 &\hfil 8.12     &\hfil 5.44          &\hfil 16.66  &\hfil 9.63  &\hfil 4.96                   &\hfil 4.44         &\hfil    4.4        &\hfil \textbf{4.06} \\ \cline{2-10} 
                        &\hfil 15 &\hfil 5.45     &\hfil 4.54          &\hfil 11.25  &\hfil 8.48  &\hfil 4.19                    &\hfil 3.73        &\hfil       3.66     &\hfil \textbf{3.28} \\ \cline{2-10} 
                        &\hfil 20 &\hfil 4.37     &\hfil 4.06          &\hfil 8.99  &\hfil 7.99  &\hfil 3.85                    &\hfil 3.36        &\hfil      3.43      &\hfil \textbf{2.99} \\ \hline
\multirow{5}{*}{Music}  &\hfil 0  &\hfil 23.4     &\hfil 9.62          &\hfil 16.24  &\hfil 12.92  &\hfil 8.94                   &\hfil 7.79         &\hfil    8.03        &\hfil \textbf{7.35} \\ \cline{2-10} 
                        &\hfil 5  &\hfil 13.5     &\hfil 6.6          &\hfil 11.44  &\hfil 10.1  &\hfil 5.83                    &\hfil 5.23        &\hfil     5.49       &\hfil \textbf{4.9}  \\ \cline{2-10} 
                        &\hfil 10 &\hfil 7.96     &\hfil 5.05          &\hfil 9.13  &\hfil 8.95     &\hfil 4.76                 &\hfil 4.11           &\hfil       4.19     &\hfil \textbf{3.69} \\ \cline{2-10} 
                        &\hfil 15 &\hfil 5.54     &\hfil 4.47          &\hfil 8.10  &\hfil 8.35   &\hfil 4.17                   &\hfil 3.63         &\hfil     3.78       &\hfil \textbf{3.14} \\ \cline{2-10} 
                        &\hfil 20 &\hfil 4.55     &\hfil 4.06          &\hfil 7.48  &\hfil 7.95   &\hfil 3.72                   &\hfil 3.30         &\hfil     3.47       &\hfil \textbf{2.93} \\ \hline
\multirow{5}{*}{Noise}  &\hfil 0  &\hfil 21.41    &\hfil 9.26          &\hfil 16.56  &\hfil 13.12   &\hfil 8.09                  &\hfil 7.34          &\hfil   7.68         &\hfil \textbf{6.8}  \\ \cline{2-10} 
                        &\hfil 5  &\hfil 13.43    &\hfil 6.81          &\hfil 12.26  &\hfil 10.57    &\hfil 6.37                 &\hfil 5.65           &\hfil      5.85      &\hfil \textbf{5.23} \\ \cline{2-10} 
                        &\hfil 10 &\hfil 8.64     &\hfil 5.42          &\hfil 9.86     &\hfil 9.28   &\hfil 5                   &\hfil 4.35            &\hfil      4.6      &\hfil \textbf{4.07} \\ \cline{2-10} 
                        &\hfil 15 &\hfil 6.3      &\hfil 4.51          &\hfil 8.69  &\hfil 8.59   &\hfil 4.29                   &\hfil 3.85         &\hfil     3.9       &\hfil \textbf{3.39} \\ \cline{2-10} 
                        &\hfil 20 &\hfil 4.94     &\hfil 4.22          &\hfil 7.83  &\hfil 8.1   &\hfil 3.99                    &\hfil 3.44        &\hfil      3.74      &\hfil \textbf{3.1}  \\ \Xhline{2\arrayrulewidth}
\multicolumn{2}{c|}{Average (EER / $C_{det}^{min}$)} &\hfil 10.85 / 0.758    &\hfil 6.12 / 0.31          &\hfil 13.52 / -  &\hfil 10.24 / -  &\hfil 5.6 / 0.315 &\hfil 5.07 / 0.563 &\hfil 5.15 / 0.286&\hfil \textbf{4.55} / \textbf{0.254} \\ \Xhline{2\arrayrulewidth}
\end{tabular}
}
\vspace{-0.2cm}
\end{table*}

\subsection{ExU-Net}
Although the described U-Net-based system can induce discriminative and noise-mitigated embeddings, the enhanced spectrograms extracted from the decoder are not directly exploited for speaker embedding extraction. 
In other words, the indirect utilization of the decoder for noise compensation training suggests that all parameters of the trained U-Net cannot be fully used for embedding extraction. 
Therefore, we propose an extended U-Net (ExU-Net) using an additional extractor to alleviate these structural weaknesses. 

The solid blue arrows in Figure \ref{fig:overall} depict the feedforward flow of ExU-Net. 
The extractor is identical to the encoder structure, except for the concatenation layer added at the beginning of each block. 
The speaker embeddings ($B$, $\tilde{B} \in \mathbb{R}^{256}$) of the proposed ExU-Net are derived from the extractor network, $f_{ext}$, by directly feeding the enhanced spectrogram from the decoder as follows: 
\begin{equation}
\vspace{-0.05cm}
\begin{aligned}
    B = g(f_{ext}(O)) =  g(f_{ext}(f_{dec}(f_{enc}(S)))) , \\ 
    \tilde{B} = g(f_{ext}(\tilde{O})) = g(f_{ext}(f_{dec}(f_{enc}(\tilde{S})))),  
\end{aligned}
\end{equation}
Thus, the ExU-Net's SID loss ($\mathcal{L'}_{CCE}$) optimizes the speaker embeddings extracted from the enhanced spectrograms, in contrast to Eqs. (\ref{eq2}) and (\ref{eq3}). 

Additionally, considering the characteristics of the SV task comparing the similarity between embeddings, we applied metric learning to directly enhance the embeddings. 
For the EE loss, we used the angular prototypical network loss (APN) \cite{chung2020defence}. 
\begin{equation}
T_{i,j} =  w \cdot cos({B}_{i}^{1}, \tilde{B}_{j}^{2}) + b, 
\end{equation}
where $T_{i,j}$ is the cosine similarity ($cos$) between ${B}_{i}^{1}$ and $\tilde{B}_{j}^{2}$ embeddings with the learnable weight and bias $w$ and $b$, respectively. 
The EE loss ($\mathcal{L}_{APN}$) encourages to explore noise-robust and discriminative embedding spaces by decreasing the intra-class variance between noise-clean embedding pairs of the same speaker and increasing the inter-class variance between different speakers. 
\begin{equation}
\mathcal{L}_{APN} = -\frac{1}{n}\sum_{j=1}^{n}log\frac{exp(T_{j,j})}{\sum_{i=1}^{n}exp(T_{i,j})},
\end{equation}
Finally, the ExU-Net is trained jointly to optimize the following three losses: 
\begin{equation}
    \mathcal{L}_{ExUnet} = \mathcal{L'}_{CCE} + \mathcal{L}_{MSE} + \mathcal{L}_{APN}.
\end{equation}

\section{Experimental Setting}
\label{sec:experimental setting}

\subsection{Datasets}
Experiments were performed on the VoxCeleb1 \cite{voxceleb} and VOiCES \cite{richey18_interspeech} datasets. 
VoxCeleb1 consists of a training set with 148,642 utterances from 1,211 speakers and a test set with 4,874 utterances from 40 speakers. 
Although the VoxCeleb dataset collected from YouTube videos is moderately noisy, we considered the original utterances clean and generated the noise data additionally. 
The MUSAN corpus \cite{snyder2015musan} was used for noise sources, and we divided the MUSAN dataset into two nonoverlapping parts for data augmentation in the training and test phases. 
In training, we used the original VoxCeleb1 training dataset $D$ alone or with the noise set ($D^N$) synthesized using the MUSAN training subset with a signal-to-noise ratio (SNR) randomly selected between 0 and 20. 

For the evaluation under noise scenarios, we used two test sets. 
First, for the VoxCeleb1 test set, we used the synthesized test data for each noise type with SNRs in the set \ \{0, 5, 10, 15, 20\} using the MUSAN test subset and the original test set $\tau$. 
Second, we exploited the VOiCES development dataset of 15,904 audio segments from 196 speakers for different domain noise scenarios. 
The VOiCES dataset was collected using array microphones at diverse distances and acoustic conditions in rooms of various sizes. 
We compared the models based on the equal error rate (EER) and minimum detection cost function ($C_{det}^{min}$) using the cosine similarity score, as in \cite{chung2018voxceleb2}.

\subsection{Implementation Details}
We employed a 64-dimensional mel-spectrogram as input extracted with a 1024-point FFT and a Hamming window width of 25 ms, hopped at 10 ms. 
The mini-batch consisted of one clean and one noisy utterance per 60 randomly selected speakers, totaling 120 utterances. 
In all experiments, we used the Adam optimizer with a learning rate of 0.001. 
Each model was trained for 500 epochs, and the learning rate was decreased by 5\% every 10 epochs. 
The batch normalization \cite{ioffe2015batch} and rectified linear unit activation \cite{nair2010rectified} functions were applied after convolutional layers. 
This study did not use any voice activity detection technique. 
For reproducibility, we provide the experimental codes and weights of the trained model at \url{https://github.com/wngh1187/ExU-Net}. 

\begin{table}[]
\caption{
    Ablation experiments of ExU-Net loss functions. 
}
\vspace{-0.1cm}
\centering
\label{table:ablation_loss}
\begin{tabularx}{\linewidth}{l|C C C|C C C C}
\Xhline{2\arrayrulewidth}
\multirow{2}{*}{Systems} &\multicolumn{3}{c|}{Loss functions} & \multicolumn{4}{c}{Vox1 (EER \% / $C_{det}^{min}$)} \\
\cline{2-8}
 & SID & FE & EE & \multicolumn{2}{c}{Original} & \multicolumn{2}{c}{Noises mean} \\ \hline
\#1 & $\times$ & MSE & APN & \multicolumn{2}{c}{3.62 / 0.25} & \multicolumn{2}{c}{5.66 / 0.322} \\
\#2 & CCE & $\times$ & APN & \multicolumn{2}{c}{2.82 / 0.18} & \multicolumn{2}{c}{4.83 / 0.282} \\ 
\#3 & CCE & MSE & $\times$ & \multicolumn{2}{c}{3.32 / 0.227} & \multicolumn{2}{c}{5.38 / 0.318} \\
\#4 &CCE & MSE & MSE & \multicolumn{2}{c}{3.09 / 0.201} & \multicolumn{2}{c}{4.75 / 0.272} \\
\#5 & CCE & MSE & APN & \multicolumn{2}{c}{\textbf{2.76} / \textbf{0.178}}  & \multicolumn{2}{c}{\textbf{4.67} / \textbf{0.26}} \\
\Xhline{2\arrayrulewidth}
\end{tabularx}
\vspace{-0.2cm}
\end{table}

\section{Results}
\label{sec:results}
In Table \ref{table:sota}, we compare the models with the recently proposed SV systems for noisy environments on the VoxCeleb test set. 
For the convenience of description, we calculated the average of the EER and $C_{det}^{min}$ derived from all evaluation conditions. 
The baseline system trained only on the original clean data $D$ has difficulty coping with noise utterances, resulting in significant degradation under noisy scenarios. 
When comparing the results of the baselines under the two training dataset conditions, simply using augmented training data $D^N$ could enhance the robustness to noise utterances (6.12\% vs. 10.85\%). 
In addition, the proposed U-Net-based system (U-Net) showed a relative error reduction (RER) of 8.5\% based on the average EER compared to the baseline trained using $D + D^N$ (5.6\% vs. 6.12\%). 
The ExU-Net proposed in this study outperformed other models based on the original test trial and achieved state-of-the-art performance in all noise scenarios, with a 25.7\% of RER vs. the baseline (4.55\% vs. 6.12\%). 
Furthermore, to demonstrate the superiority of the ExU-Net structure, we additionally constructed a lightweight variant of the ExU-Net (ExU-Net-L) by adjusting the number of parameters similar to the baseline. 
The ExU-Net-L demonstrated better results than the baseline and even the U-Net and exhibited tolerable performance compared to the recently proposed models. 
Based on these results, we interpreted that the fully joint training scheme is effective for noisy SV tasks, and the proposed ExU-Net is further robust to noisy environments, including clean scenarios. 

Table \ref{table:ablation_loss} presents the results of the ablation experiments to analyze the effectiveness and appropriateness of each ExU-Net loss function. 
The model was evaluated based on the EER and $C_{det}^{min}$ of the original test set and the mean of all noise-type evaluations. 
Systems 1 and 3 exhibited notable performance degradation compared to the ExU-Net (System 5), whereas System 2 had comparable results even though FE learning was excluded. 
These results indicate that the SID loss is crucial to the generalization of the proposed model, and noise compensation at the embedding level is more effective than at the feature level. 
System 4 exhibited greater deterioration in the original evaluation conditions than the noise test sets compared to System 5. 
These results imply that the MSE used for EE learning can compensate for noise utterances by reducing the Euclidean distance between pairs of clean-noise embeddings, but it cannot consider the inter-class variance between speakers as opposed to APN loss. 

Table \ref{table:voices} delivers the results of the VOiCES development set for each model. 
All models were trained using the VoxCeleb1 training set ($D+D^N$). 
Similar to the previous results, the U-Net displayed improved performance compared with the baseline. 
In addition, ExU-Net achieved the best performance among all models and exhibited an RER of 15.3\% vs. the baseline. 
Through the results for the evaluation dataset in different domains from the training set, the models demonstrated superior generalization. 

\begin{table}[]
\caption{Results of EER and $C_{det}^{min}$ on the VOiCES development set.}
\vspace{-0.1cm}
\centering
\label{table:voices}
\resizebox{\linewidth}{!}{%
\begin{tabular}{c | m{1.1cm} m{1cm} m{1.2cm} m{1cm} m{1.1cm}}
\Xhline{2\arrayrulewidth}
&\hfil Baseline &\hfil VoiceID &\hfil Wu \textit{et. al.} &\hfil U-Net & ExU-Net       \\ \hline
EER &\hfil 7.71 &\hfil 13.57 &\hfil 12.51 &\hfil 6.97 &\hfil \textbf{6.53} \\ 
$C_{det}^{min}$ &\hfil 0.44 &\hfil - &\hfil - &\hfil 0.42 &\hfil \textbf{0.38} \\ 
\Xhline{2\arrayrulewidth}
\end{tabular}
}
\vspace{-0.2cm}
\end{table}

\section{Conclusion}
\label{sec:conclusion}
Focusing on the drawbacks of the independent enhancement module for noise utterance in SV tasks, we emphasized the necessity of a fully joint training scheme between noise compensation and SV models. 
This paper proposed models that simultaneously optimize feature (and embedding) enhancement and SID losses using the U-Net or extending it (ExU-Net). 
Due to the joint learning of two or three tasks, the proposed models can extract noise-robust speaker embeddings customized for SV tasks. 
The evaluation of the noise-synthesized VoxCeleb1 test set demonstrated that the U-Net-based system exhibited improved performance compared to the baseline, and ExU-Net achieved state-of-the-art performance. 
In addition, we proved the effectiveness and validity of the ExU-Net loss functions through ablation experiments. 
Finally, the proposed models demonstrated outstanding generalization performance in the evaluation of the VOiCES dataset. 
As a future work, we will evaluate the proposed models in various real-world noise environments. 

\vspace{-0.1cm}
\section{Acknowledgement}
This research was supported by Basic Science Research Program through the National Research Foundation of Korea (NRF) funded by the Ministry of Science, ICT \& Future Planning (2020R1A2C1007081).

\bibliographystyle{IEEEtran}

\bibliography{mybib}

\end{document}